\documentclass{article}

\usepackage{microtype}
\usepackage{graphicx}
\usepackage{subfigure}
\usepackage{booktabs} 
\usepackage{bm}
\usepackage{amssymb}
\usepackage{tikz}
\usepackage{natbib}
\usepackage{graphicx}
\usepackage{amsmath}
\usepackage{enumitem}
\usepackage{dblfloatfix}
\UseRawInputEncoding
\usepackage[hyphens]{url}
\usepackage{hyperref}
\usepackage[hyphenbreaks]{breakurl}

\usepackage[accepted]{icml2021}

\icmltitlerunning{Secure Bayesian Federated Analytics for Privacy-Preserving Trend Detection}

\begin{document}

\twocolumn[
\icmltitle{Secure Bayesian Federated Analytics for Privacy-Preserving Trend Detection}



\icmlsetsymbol{equal}{*}

\begin{icmlauthorlist}
\icmlauthor{Amit Chaulwar}{to}
\icmlauthor{Michael Huth}{to,goo}
\end{icmlauthorlist}

\icmlaffiliation{to}{Xayn AG, Berlin, Germany}
\icmlaffiliation{goo}{Department of Computing, Imperial College London, London, U.K.}

\icmlcorrespondingauthor{Amit Chaulwar}{amit.chaulwar@xayn.com}
\icmlcorrespondingauthor{Michael Huth}{michael@xayn.com}

\icmlkeywords{Federated Analytics, Trend Detection, Secure Aggregation}

\vskip 0.3in
]



\printAffiliationsAndNotice{}  

\begin{abstract}
Federated analytics has many applications in edge computing, its use can lead to better decision making for service provision, product development, and user experience. We propose a Bayesian approach to trend detection in which the probability of a keyword being trendy, given a dataset, is computed via Bayes' Theorem; the probability of a dataset, given that a keyword is trendy, is computed through secure aggregation of such conditional probabilities over local datasets of users. We propose a protocol, named SAFE, for Bayesian federated analytics that offers sufficient privacy for production-grade use cases and reduces the computational burden of users and an aggregator. We illustrate this approach with a trend detection experiment and discuss how this approach could be extended further to make it production-ready.
\end{abstract}

\section{Introduction}
\label{Introduction}
Access to high-quality data for data-driven decision making is very critical. With rising digitization, it has become easier to gather data. Nowadays, many human activities are digitized and we see an exponential increase in the number of IoT devices. The data generated by such devices helps with creating better products and services for the people. However, such detailed data is often of a personal nature and so poses a threat to privacy. Therefore, access to such data, even with good intentions, needs to be controlled.

Federated learning \cite{7447103, DBLP:conf/ccs/BonawitzIKMMPRS17} is the methodology that provides a means of decentralized computations for machine learning without a need for moving local data of users. In each round of the federated learning, the participating devices train a model on their respective local data and send only an encrypted update to the aggregator.  The aggregator combines updates from all participants to improve a shared model followed by its distribution to all participants. 

Federated learning has the potential of generating better models with lower latency and power consumption while also ensuring privacy. However, as there is no access to actual data from participating devices, it poses a problem for the analysis of federated learning models. Federated analytics \cite{FedAnalytics} is a practice introduced to solve this problem. It uses the same infrastructure as federated learning to aggregate the computed metric by each participating device using local data and shared models. 

Federated analytics has already gone beyond just measuring the quality metric to computing descriptive statistics \cite{FedAnalytics, zhu2020federated}, generating synthetic data \cite{9054559, chaulwar2020private} and learning new insights \cite{DBLP:journals/corr/abs-1903-10635}. These methods are generally combined with secure aggregation protocols to ensure strong privacy properties. 

This paper introduces a new topic in  federated analytics, namely \emph{Bayesian Federated Analytics}, which learns individual distributions of data on the edge and aggregates them in a privacy-preserving  way,  such  that  the  aggregated  distribution leads to new insights about the whole population. We propose a secure aggregation protocol for aggregating the local discrete distributions along with their privacy and security analysis. We describe the application of this approach primarily for a trend detection application. However, we think the scope for Bayesian federated analytics is much wider and we hope that this work will inspire new privacy-preserving approaches in Bayesian federated analytics.

{\bf Outline of paper:} Section \ref{SecBackground} describes the previously proposed methodologies for federated analytics and trend detection along with their limitations. Section \ref{Sec:BayFedAnalytics} introduces Bayesian federated analytics and discusses possible advantages over existing federated analytics. Section \ref{Sec:SecureAgg} proposes a protocol for secure aggregation of feature vectors, called SAFE in short. Its application in privacy-sensitive trend detection is given in Section \ref{Sec:TrendDetection}. Preliminary experimental results for this trend detection method are presented with a simple example in Section \ref{Sec:Results}. Finally, Section \ref{Sec:Conclusion} concludes and scopes out future work needed for making this approach production-ready.

{\bf Notation:} Throughout this paper, vectors are denoted by lower case bold letters and random variables are represented by sans serif font.

\section{Background} \label{SecBackground}
This section reviews previous approaches for federated analytics and trend detections.

\subsection{Federated Analytics}
Federated analytics originated with the need to measure the quality of the models obtained by federated learning. It uses the same infrastructure as federated learning but without the learning part. Therefore, it inherits some of the benefits as well as challenges associated with federated learning. It allows analysis of distributed data while preserving privacy. However, it also faces issues like limited communication bandwidth, exposure to adversarial attacks, and so forth.

Sharemind \cite{10.1007/978-3-540-88313-5_13}, RAPPOR \cite{Erlingsson_2014}, PROCHLO \cite{Bittau_2017} and Secure MPC for analytics \cite{7839796} are some of the privacy-preserving analytics systems that use different privacy-preserving techniques such as privacy-preserving randomization, local differential privacy, and secure multi-party computation. However, all of these methods require an intermediary between the data owners and analysers. This not only increases the computation and communication requirements but also restricts the usage of data to limited, authorised analysers. This is not desirable for the progress of research in fields like machine learning where the ease of access to the data is one of the main reasons for progress.

In the secure aggregation of a song-recognition protocol, described in  \cite{FedAnalytics}, the aggregator can see only masked values. The stronger privacy properties are ensured only with a large number of users and using local differential privacy. This does not guarantee strong privacy for a small number of users.

Federated heavy-hitters discovery \cite{zhu2020federated} is the latest privacy-preserving method that can be used to detect heavy-hitters. However, it is not suitable for finding new trends. This method has inherent initial bias as it initially samples a random subset of users. Even after the removal of stop-words, it will still detect~--~with high probability~--~heavy hitters that are not important for trend detection. Also, it needs to update its belief for trend identification based on past knowledge. Otherwise, the same heavy hitters may be detected as a trend in consecutive iterations. Since a single heavy-hitter discovery requires multiple rounds, this is not efficient as well. 

\subsection{Trend Detection}
The most common way of quantifying a trend is to look at the counts of keywords, mentions or other engagements over a period of time. Atypical changes in such time series can indicate the presence of a trend, but a general definition of “atypical” is hard to develop and largely depends on the specific needs. The factors influencing the detection of trends include time-to-detection, robustness, precision, recall, etc. A survey of trend detection methods can be found in \cite{instance1290, SwarajK2018RecentAF}.

The most obvious application of trend detection is social media trend detection. The Twitter trend detection methods are described in \cite{instance1290}. However, a research study \cite{Annamoradnejad_2019}, has shown that 68\% of Twitter trends are hashtags. Therefore, the usage of these methods for trend detection in a set of documents without hashtags is not evaluated. Facebook trend detection \cite{6118886} performs descriptive statistical analysis to detect three different categories of trending topics, which they call disruptive events, popular topics, and daily routines. As the data for trend detection based on Twitter and Facebook is generally public, their trend detection methods do not need to consider the privacy of users.

Google Trends \cite{GoogleTrends} is a website by Google that analyzes the popularity of top search queries in Google Search across various regions and languages. Although it only publishes trending search queries, Google itself has  access to each user's individual search history which is very sensitive information. 

The information of which webpages a user browsed will be even more insightful and very privacy-sensitive. Therefore, there are benefits in creating a trend detection method for  browsed webpages that is secure and does not leak any private information.

\section{Bayesian Federated Analytics} \label{Sec:BayFedAnalytics}
We take an alternative Bayesian view when compared to traditional federated analytics methods of measuring counts or rates. We now present this approach and describe its benefits that can be exploited using a Bayesian perspective. We note that the original motivation for applying Bayesian thinking to federated analytics is taken from \cite{WellingMax}. 

The basic idea behind the Bayesian federated analytics is to learn individual data distributions of users at the edge and to aggregate them securely such that the aggregated distribution generates insights that help with making better decisions, products, and so forth. We believe that such a Bayesian approach for federated analytics has the following potential advantages:
\begin{itemize}
    \item The distribution of the edge data is determined by a few parameters. This reduces the necessary communication bandwidth for transferring these local parameters for computing those of the aggregated distribution.
    
    \item Distributions approximate raw local data. Therefore, their parameters contain less information than the raw data and so their communication is less risky than communicating raw data.
    
    \item Representing local data via distributions makes the application of privacy-preserving techniques such as differential privacy easier. 
    
    \item The Bayesian approach allows for belief updates by considering the posteriors from the last step as a prior for the current step. In contrast, the aggregated statistics of traditional federated analytics needs to be tracked and additional time series analysis is required to learn new trends.
    
    \item In the Bayesian approach, an active adversary can only influence the aggregation results within the value ranges of distribution parameters. Violations of such ranges can be detected and such aggregation input may be refused. 
         
    \item The aggregated distributions can be used to generate synthetic data, for example for data augmentation~--~using generative models like variational autoencoders, and generative adversarial networks.  
\end{itemize}

\section{Secure Aggregation of Feature Vectors} \label{Sec:SecureAgg}
This section proposes a protocol, SAFE, for the secure aggregation of feature vectors of $N>1$ users in an honest but curious adversary model. The set of features for all participating users is indicated as $\mathbb{V}$. For example, this work takes a set of English vocabulary of size $d$ as feature set. 

The $d$-dimensional \emph{feature vector} $\bm v_{i} = (v_{i,1},\dots, v_{i,d})$ describes the value $\bm v_{i,j}$ for all features $j$ of user $i$. We assume that all feature values $\bm v_{i,j}$ are in the closed interval $[a^j,b^j]$ for all users $i$. For our application, we have $[a^j,b^j] = [0,1]$ for all $j=1,\dots, d$ as feature values are  probabilities. The target function for aggregations is $\sum_{i=1}^N \bm v_i$, $\sum_{i=1}^N \bm v_i/N$, or similar variants. 

In this protocol, the feature values will be obfuscated with sums of shares, where \emph{randomly generated} shares  are from an interval of doubles. For example, this could be $[-D, D]$ for some large double $D > 0$. The choice of this interval may be a function of the round number $r$ of the aggregation protocol, the user $i$, the coordinate $\bm v_{i,j}$ of the vector $\bm v_i$ or additional factors. The steps of this protocol in a single round of aggregation are as follow:

\begin{enumerate}
    \item Each user $i$ creates $N$ shares (where $N$ is the total number of users) for her feature vector. The steps for share creation and distribution for feature vector $\bm v_{i}$ are:
    \begin{enumerate}
        \item $N-1$ shares $\tilde{\bm v}^k_{i}$ are randomly and independently drawn in $[-D, D]^d$, where $k$ ranges over $\{1,2,\ldots,N\}\setminus \{i\}$.
        
        \item The $N^{th}$ share $\tilde{\bm v}^i_i$ is calculated as follows:
        \begin{equation} \label{EqSharesCreation}
       \tilde{\bm v}^i_{i} = \bm v_{i} - \sum_{k\not=i} \tilde{\bm v}^k_{i}.
    \end{equation}
    
    As the $N-1$ shares are calculated randomly from random intervals, the $N^{th}$ share $\tilde{\bm v}^i_{i}$ is also not deterministic.  
    
    \end{enumerate}
    \item Each user $i$ keeps her $N^{th}$ share $\tilde{\bm v}_i^i$. For each $k$ in $\{ 1,2,\dots, N\} \setminus \{i\}$, user $i$ sends the share $\tilde{\bm v}_{i}^{k}$ to the user $k$.

    In the same manner, user $i$ also receives $N-1$ shares from the other users: from each $k$ in  $\{ 1,2,\dots, N\}\setminus \{i\}$, user $i$ receives $\tilde{\bm v}_{k}^i$.
    \item User $i$ can now calculate the obfuscated feature vector $\bm v'_i$ for user $i$ as:
    \begin{equation} \label{EqUpdateFeatureVector}
        \bm v'_i = \tilde {\bm v}_i^i + \sum_{k\not=i} \tilde{\bm v}_{k}^i.
    \end{equation}
     \item Each user $i$ sends the obfuscated feature vector $\bm v'_i$ to the aggregator.
     \item The aggregator computes $\sum_{i=1}^N \bm v'_i$ as the target output since the following relation holds.
     \begin{equation}\label{equ:key}
         \sum_{1\leq i\leq N} \bm v_i = \sum_{1\leq i\leq N} \bm v'_i.
     \end{equation}
     To see that this is correct, we first note that we have
\begin{equation}
{\bm v}_i = \sum_{k=1}^N \tilde {\bm v}_i^k    
\end{equation}

\noindent for all users $i$. Therefore, the sum of all shares of all users equals the lefthand side of~(\ref{equ:key}). Thus, it suffices to show that the righthand side in~(\ref{equ:key}) is the sum of all shares of all users. But this is so since the diagonal shares $\tilde {\bm v}_i^i$ occur in the sum of shares ${\bm v}_i'$ for user $i$. The off-diagonal shares $\tilde {\bm v}_i^k$ of user $i$ (where $k\not=i$) occur in the sum ${\bm v}_k'$. This proves that~(\ref{equ:key}) holds.

\end{enumerate}

\tikzstyle{block} = [rectangle, draw, fill=blue!20, 
    text width=5em, text centered, rounded corners, minimum height=5em]
\tikzstyle{block1} = [rectangle, draw,text width=5em, text centered, rounded corners, minimum height=3em]

\newcommand*{\connectorH}[4][]{
  \draw[#1] (#3) -| ($(#3) !#2! (#4)$) |- (#4);
}
\newcommand*{\connectorV}[4][]{
  \draw[#1] (#3) |- ($(#3) !#2! (#4)$) -| (#4);
}

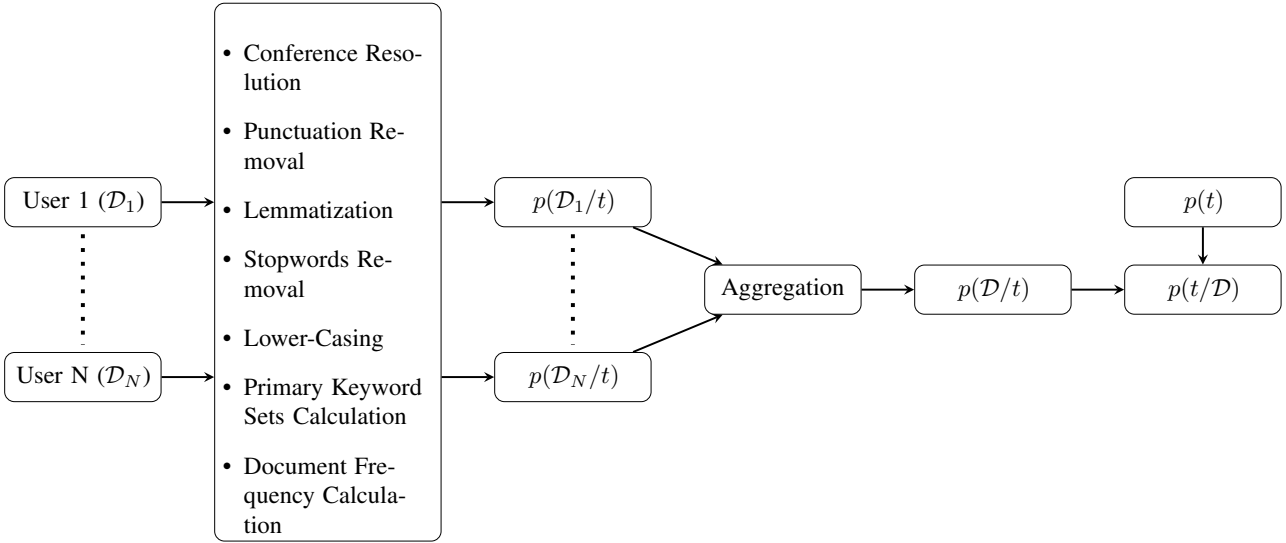
\begin{figure*}
    \centering
     \resizebox{\textwidth}{!}
    {
    \begin{tikzpicture}
     \node (user1) [block1, text width=2 cm,align=center, minimum height=2em] {User 1 ($\mathcal{D}_1$)};
     \node (userN) [block1, below of = user1, yshift = -1.5 cm, text width=2 cm,align=center, minimum height=2em] {User N ($\mathcal{D}_N$)};
     \node (preprocess) [block1, right of =user1, xshift=2.5 cm, yshift=-1 cm, text width=3 cm,align=center, minimum height=2em] {\begin{itemize}[leftmargin=*]
         \item Conference Resolution
         \item Punctuation Removal
         \item Lemmatization
         \item Stopwords Removal
         \item Lower-Casing
         \item Primary Keyword Sets Calculation
         \item Document Frequency Calculation
     \end{itemize}};

     \node (userN_likelihood) [block1, right of = userN,text width=2 cm,align=center, xshift = 6 cm, minimum height=2em] { $p(\mathcal{D}_N/t)$};
     \node (user1_likelihood) [block1, right of = user1,text width=2 cm,align=center, xshift = 6 cm,  minimum height=2em] { $p(\mathcal{D}_1/t)$};
     \node (likelihood_aggregation) [block1, right of = userN_likelihood,text width=2 cm,align=center, xshift = 2 cm, yshift = 1.25cm, minimum height=2em] { Aggregation};
     \node (likelihood_aggregated) [block1, right of = likelihood_aggregation,text width=2 cm,align=center, xshift = 2 cm,   minimum height=2em] {  $p(\mathcal{D}/t)$};
     \node (trend)[block1,text width=2 cm,align=center, minimum height=2em, right of = likelihood_aggregated, xshift=2 cm]{$p(t/\mathcal{D})$};
     \node (prior) [block1,text width=2 cm,align=center, minimum height=2em] at (user1_likelihood -| trend) {$p(t)$};
     
     \draw [thick,->,>=stealth] (user1.east) -- (user1-|preprocess.west);
     \draw[{ultra thick, loosely dotted, shorten >=1mm, shorten <=1mm}](user1) -- (userN);
     \draw [thick,->,>=stealth] (userN.east) -- (userN-|preprocess.west);
     \draw [thick,->,>=stealth] (preprocess.east |- user1_likelihood) -- (user1_likelihood);
     \draw [thick,->,>=stealth] (preprocess.east |- userN_likelihood.west) -- (userN_likelihood);
     \draw[{ultra thick, loosely dotted, shorten >=1mm, shorten <=1mm}](user1_likelihood) -- (userN_likelihood);
    \draw [thick,->,>=stealth] (user1_likelihood) -- (likelihood_aggregation);
    \draw [thick,->,>=stealth] (userN_likelihood) -- (likelihood_aggregation);
    \draw [thick,->,>=stealth] (likelihood_aggregation) -- (likelihood_aggregated);
    \draw [thick,->,>=stealth] (likelihood_aggregated) -- (trend);
    \draw [thick,->,>=stealth] (prior) -- (trend);
    \end{tikzpicture}
    }
    \caption{Flow Diagram for Computing the Probabilty of a Keyword $t$ to be a Trending Keyword with Local Data $\mathcal{D}_1, \ldots, \mathcal{D}_N$ }
    \label{fig:TrendDetection}
\end{figure*}

We note that users receive random vectors as shares from other users. Only $\tilde {\bm v}_i'$ may leak some information about $\bm v_i$ and only the aggregator will receive this private input. But this leakage is considerably mitigated against, since $N-1$ random shares from $N-1$ users are combined with the $N^{th}$ share of user $i$ to compute this private input. 

More secure solutions require the exchange of secrets between users, mapping feature values into other representations for perfectly secure encryption (e.g.\ modulo arithmetic), or the use of homomorphic encryption. Our protocol offers sufficient protection for trend detection in practical settings while computing an aggregate with minimal effort.

\section{Bayesian Federated Trend Detection} \label{Sec:TrendDetection}
In this section, we apply protocol SAFE to privacy-preserving detection of trends for document interactions (e.g., webpages clicked/liked) based on our approach to Bayesian federated analytics. 

We define the trend for interacted documents as below:

\begin{quote} \label{Quote:TrendDefinition}
    \emph{{\bf Trend:} Keywords that are not frequently present in past interacted documents but that appear frequently in current interacted documents. }
\end{quote}

 This definition is very basic and neglects many factors such as seasonality. Our aim here is to illustrate our Bayesian approach to federated and privacy-preserving analytics.

We formulate the task as the determination of the trending keyword $t$ in $\mathbb{V}$ while $N$ users have interacted with random subset of document set $\mathcal{D}$. We note that multiple users can interact with same document.  The term $\mathbb{V}$ represents a set of all possible keywords that can be trending. The set of documents, with which user $i$ has  interacted with, is indicated as $\mathcal{D}_i$ and so $\mathcal{D}_i \subset \mathcal{D}$. In probabilistic terms, the task is to find a distribution $p({\mathrm{t}}\mid \mathcal{D})$ that defines a higher probability for a keyword in $\mathbb{V}$ as per our definition of the trending keywords.

According to the Bayes formula, the expression for this term can be written as
\begin{equation} \label{EqTrendBayes}
    p({\mathrm{t}=t}\mid \mathcal{D}) = \frac{p(\mathcal{D}\mid {\mathrm{t}}=t)p({\mathrm{t}}=t)}{p(\mathcal{D})}.
\end{equation}

This equation also captures our above definition of trend. 
The semantics and its computation for each term from the right hand side of Eq.\,(\ref{EqTrendBayes}) are described in following subsections and depicted in Fig.\,\ref{fig:TrendDetection}.

\subsection{Prior Calculation}
The term $p({\mathrm{t}})$ is the prior discrete probability distribution over all keywords in the vocabulary $\mathbb{V}$ for being a trending keyword. As per the definition of trend, the keyword which occurs frequently in previously interacted documents should have a low prior probability for being a trending keyword. The words which occur frequently in the past interacted documents have low inverse document frequency (IDF) values. 

Therefore, the prior distribution is modelled as a Dirichlet distribution parametrized by the IDF of keywords and the prior probability for each keyword is calculated as the expected mean from the Dirichlet distribution.  As past interacted documents, we take the whole English Wikipedia to demonstrate the methodology. However, any suitable dataset can be chosen for calculating the prior probabilities. If there is no available dataset, one can also start with uniform prior and update the prior over the iterations.

\begin{table*}[!b]
\caption{Trend Detection Results with 10 Virtual Users}
\vspace{-0.3 cm}
\label{Table:TrendDetectionMSMArco}
\vskip 0.15in
\begin{center}
\begin{small}
\begin{sc}
\begin{tabular}{lccccc}
\toprule
 Keyword &  Total Count in $\mathcal{D}$ &  IDF & \shortstack{Ranking \\ based on IDF}  & \shortstack{Ranking \\ based on Total Count} & \shortstack{Ranking \\ based on Pooled Trend}    \\
\midrule
phloem    & 49 & 9.8125 & 1& 7  & 3 \\
xylem    &  30  & 9.6191  & 2 & 16 & 4 \\
offender & 48  & 7.3567 & 3 & 8 & 3\\
rica    & 57  & 6.0512 & 4 & 5 & 1        \\
costa     & 57  & 5.2358 & 5 & 4 & 1\\
manhattan & 77  & 5.14365 & 6 & 1 & 2\\
project & 75  & 3.1363 & 12 & 2 & 2\\
\bottomrule
\end{tabular}
\end{sc}
\end{small}
\end{center}
\vskip -0.1in
\end{table*}

\subsection{Likelihood Calculation}
 The term $p(\mathcal{D}\mid{\mathrm{t}})$ is the likelihood for the Document set $\mathcal{D}$ given that the keyword ${\mathrm{t}}$ is trending. As the aim is to protect the privacy of each user, nobody but user $i$ has access to the document set $\mathcal{D}_i$. Therefore, each user calculates individual likelihoods $p(\mathcal{D}_i|\mathrm{t})$ which then are aggregated in a privacy-preserving way to the likelihood of the whole document set. 

The likelihood $p(\mathcal{D}_i|{\mathrm{t}})$ is calculated based on the number of documents in which the keyword is present in the primary keyword set of that document. The primary keyword set is the set of five keywords that have the highest term frequency in a document.  We now describe these steps in detail:

Using the conference resolution NLP task \cite{clark2016deep}, all the pronouns are converted to corresponding entities. We do this since important keywords in a document are mentioned only a few times although referred to many times with pronouns. Non-conversion can adversely affect the term frequency of such words and they would not be included in the primary keyword set. We used the conference resolution model from the NLP library Spacy \cite{spacy}.


In order to reduce the computations and noise, general NLP preprocessing tasks such as punctuation removal, lower-casing, lemmatization and stop-words removal are performed on the document set. 
 
 For each document, the primary keyword sets are calculated by computing the term frequencies of keywords inside the documents.

The likelihoods of the dataset, given each keyword from the primary keyword set, are calculated as the Dirichlet distribution parametrized by the number of documents in which the keywords in primary keyword sets are present.

We adopted the above method for likelihood calculations for sake of simplicity. Different heuristics can be used instead, as per the requirements to calculate these likelihoods.

\subsection{Aggregation of Likelihood}
Each user creates a $d-$dimensional vector of likelihoods, where $d$ is the size of the vocabulary $\mathbb{V}$, as follow:
\begin{equation}
    \bm{\mathcal{L}}_i = \left(p(\mathcal{D}_i\mid {\mathrm{t}}=t_1),\ldots, p(\mathcal{D}_i\mid {\mathrm{t}}=t_d)\right),
\end{equation}
where the likelihoods for the document set given the keyword that is not present in any of the primary keyword set is 0. If each document set $\mathcal{D}_i$ is considered as an instance of the random variable for the subset of document set $\mathcal{D}$, then the aggregation of these instances, i.\,e., the likelihood of the whole document set, can be represented  as their mean. Therefore, the aggregator wants to perform the following operation for each keyword:

\begin{equation} 
    p(\mathcal{D}\mid {\mathrm{t}}=t_1) = \sum_{i=1}^N p(\mathcal{D}_i\mid {\mathrm{t}}=t_1).
\end{equation}

This is the target function of our privacy-preserving aggregation protocol SAFE described in Section \ref{Sec:SecureAgg}, using vectors of likelihoods as the feature vectors.

\subsection{Marginalised Probability of Dataset}
The calculation of the marginal probability $p(\mathcal{D})$ needs access to the entire dateset $\mathcal{D}$. 
The protection of user privacy demands that this dataset is either not computed in the first place or has restricted access. 
Therefore, we prefer not having to compute the marginal probability at all.
We now describe how we can avoid this.

The marginal probability would be constant and the posterior probability is proportional to the product of likelihood and prior probability. Therefore, the equation (\ref{EqTrendBayes}) can be alternatively written as follow:
\begin{equation}
  p({\mathrm{t}=t}\mid \mathcal{D}) \propto {p(\mathcal{D}\mid {\mathrm{t}}=t)p({\mathrm{t}}=t)}.
\end{equation}

As our aim to find the trending keywords, i.\,e., only the ranking of keywords, the calculation of exact value for the posterior is not necessary and the marginalisation constant $p(\mathcal{D})$ can be safely ignored.

\section{Results} \label{Sec:Results}
In this section, the proposed trend detection methodology is tested with 10 virtual users using a test document set formed using a small subset of passages (first 50, see \ref{App:MSMarco_passages}) from the MSMarco passage dataset  \cite{bajaj2018ms}. Each virtual user $i$ chooses a random number of passages from the test document set with replacement to form their document set $\mathcal{D}_i$. The IDF weights for the keywords are taken from \cite{GalkinM}. The results of performing the trend detection algorithm described in section \ref{Sec:TrendDetection} are summarized in Table \ref{Table:TrendDetectionMSMArco}. The proposed method is compared with the following two alternative methodologies: 
\begin{enumerate}
    \item \emph{Ranking based on total count:} This method ranks keywords based on their total count in the dataset $\mathcal{D}$. This is obviously not privacy-preserving and does not take into account any history.
    
    \item \emph{Ranking based on pooled trend:} This method pools trending keyword found locally by each user and then ranks them based on the number of users who report it as their local trend. The dataset $\mathcal{D}$ is not shared but local trends are, which can reveal sensitive information. This approach does not consider any history. 
\end{enumerate}

The results show that the keywords like ``phloem'', ``xylem'' have far less count than other keywords such as  ``Manhattan'' and ``project''. Nevertheless they were ranked higher because of high IDF values. On the other hand, ranking based on total count shows opposite results while ranking based on pooled trend shows comparable results which would look similar to ranking based on IDF if history is taken into consideration. In our proposed method, neither hyperparameters nor a time-series analysis are required. 

For this test, we used the keywords from an English Vocabulary for trend detection. However, this method can be easily extended from uni-gramss to n-grams if the corresponding vocabulary is used. Nonetheless, the uni-grams can also provide valuable insights into topics. The trending keywords from Table \ref{Table:TrendDetectionMSMArco} can be paired to discover popular topics. For example, the trending keywords pairs ``xylem-phloem'', ``costa-rica'' and ``manhatten-project'' indicate the topics \emph{plant tissues}, the \emph{country Costa Rica} and the \emph{world war 2 nuclear weapon project}, respectively. 

\section{Conclusion and Future Work} \label{Sec:Conclusion}
We proposed a Bayesian approach to federated analytics that offers several benefits for production-grade systems: strong enough privacy and efficiency gains over the traditional federated analytics. We demonstrated these potential advantages using a trend detection application in combination with a novel secure aggregation protocol. Our simple test experiment validated the potential of our proposed approach by comparing to two alternative approaches.  

The trend detection methodology presented in this paper can be extended to increase its utility further. 
Its pre-defined uni-gram vocabulary can be modified to an n-gram vocabulary. The ability to handle out-of-vocabulary keywords or to work without a predefined vocabulary is desired. We plan to evaluate aggregation methods that are more general than simple mean, such as a weighted mean. Weights can be defined based on different factors such as the number of documents the user interacted with, user history, and so forth. We plan to evaluate the methodology over multiple iterations with prior updates. 

We want to compare or integrate our secure aggregation protocol with other methods (e.g., privacy-preserving computation of common features via private set union for honest-but-curious users \cite{10.1007/978-3-540-72738-5_16}) and with the application of local differential privacy to feature vectors. 

The feature vectors for trend detection are huge but generally sparse. We therefore have an incentive to adjust our protocol to such sparse settings to reduce the communication bandwidth. This may use the determination of the private union of features of all users as a pre-processing step.


\bibliography{main.bib}
\bibliographystyle{icml2021}

\appendix
\section{Passages from MSMarco Dataset} \label{App:MSMarco_passages}
Following passages are used for evaluation of trend detection methodologies. The detected trending keywords reported in Table \ref{Table:TrendDetectionMSMArco} are highlighted.
\begin{enumerate}
    \item The presence of communication amid scientific minds was equally important to the success of the \textbf{Manhattan} \textbf{project} as scientific intellect was. The only cloud hanging over the impressive achievement of the atomic researchers and engineers is what their success truly meant; hundreds of thousands of innocent lives obliterated.
    \item The \textbf{Manhattan} \textbf{project} and its atomic bomb helped bring an end to World War II. Its legacy of peaceful uses of atomic energy continues to have an impact on history and science.
    \item Essay on The \textbf{Manhattan} \textbf{project} - The \textbf{Manhattan} \textbf{project} The \textbf{Manhattan} \textbf{project} was to see if making an atomic bomb possible. The success of this \textbf{project} would forever change the world forever making it known that something this powerful can be manmade.
    \item The \textbf{Manhattan} \textbf{project} was the name for a \textbf{project} conducted during World War II, to develop the first atomic bomb. It refers specifically to the period of the \textbf{project} from 194 â¦ 2-1946 under the control of the U.S. Army Corps of Engineers, under the administration of General Leslie R. Groves.
    \item	versions of each volume as well as complementary websites. The first websiteâThe \textbf{Manhattan} \textbf{project}: An Interactive Historyâis available on the Office of History and Heritage Resources website, http://www.cfo. doe.gov/me70/history. The Office of History and Heritage Resources and the National Nuclear Security
\item	The \textbf{Manhattan} \textbf{project}. This once classified photograph features the first atomic bomb â a weapon that atomic scientists had nicknamed Gadget.. The nuclear age began on July 16, 1945, when it was detonated in the New Mexico desert.
\item	Nor will it attempt to substitute for the extraordinarily rich literature on the atomic bombs and the end of World War II. This collection does not attempt to document the origins and development of the \textbf{Manhattan} \textbf{project}.
\item	\textbf{Manhattan} \textbf{project}. The \textbf{Manhattan} \textbf{project} was a research and development undertaking during World War II that produced the first nuclear weapons. It was led by the United States with the support of the United Kingdom and Canada. From 1942 to 1946, the \textbf{project} was under the direction of Major General Leslie Groves of the U.S. Army Corps of Engineers. Nuclear physicist Robert Oppenheimer was the director of the Los Alamos Laboratory that designed the actual bombs. The Army component of the \textbf{project} was designated the
\item	In June 1942, the United States Army Corps of Engineersbegan the \textbf{Manhattan} \textbf{project}- The secret name for the 2 atomic bombs.
\item	One of the main reasons Hanford was selected as a site for the \textbf{Manhattan} \textbf{project}'s B Reactor was its proximity to the Columbia River, the largest river flowing into the Pacific Ocean from the North American coast.
\item	group discussions, community boards or panels with a third party, or victim and \textbf{offender} dialogues, and requires a skilled facilitator who also has sufficient understanding of sexual assault, domestic violence, and dating violence, as well as trauma and safety issues.
\item	punishment designed to repair the damage done to the victim and community by an \textbf{offender}'s criminal act. Ex: community service, Big Brother program indeterminate sentence
\item	Tutorial: Introduction to Restorative Justice. Restorative justice is a theory of justice that emphasizes repairing the harm caused by criminal behaviour. It is best accomplished through cooperative processes that include all stakeholders. This can lead to transformation of people, relationships and communities. Practices and programs reflecting restorative purposes will respond to crime by: 1  identifying and taking steps to repair harm, 2  involving all stakeholders, and. 3  transforming the traditional relationship between communities and their governments in responding to crime.
\item	Organize volunteer community panels, boards, or committees that meet with the \textbf{offender} to discuss the incident and \textbf{offender} obligation to repair the harm to victims and community members. Facilitate the process of apologies to victims and communities. Invite local victim advocates to provide ongoing victim-awareness training for probation staff.
\item	The purpose of this paper is to point out a number of unresolved issues in the criminal justice system, present the underlying principles of restorative justice, and then to review the growing amount of empirical data on victim-\textbf{offender} mediation.
\item	Each of these types of communities the geographic community of the victim, \textbf{offender}, or crime; the community of care; and civil society may be injured by crime in different ways and degrees, but all will be affected in common ways as well: The sense of safety and confidence of their members is threatened, order within the community is threatened, and (depending on the kind of crime) common values of the community are challenged and perhaps eroded.
\item	The approach is based on a theory of justice that considers crime and wrongdoing to be an offense against an individual or community, rather than the State. Restorative justice that fosters dialogue between victim and \textbf{offender} has shown the highest rates of victim satisfaction and \textbf{offender} accountability.
\item	Inherent in many peoples understanding of the notion of ADR is the existence of a dispute between identifiable parties. Criminal justice, however, is not usually conceptualised as a dispute between victim and \textbf{offender}, but is instead seen as a matter concerning the relationship between the \textbf{offender} and the state. This raises a complex question as to whether a criminal offence can properly be described as a dispute.
\item	Criminal justice, however, is not usually conceptualised as a dispute between victim and \textbf{offender}, but is instead seen as a matter concerning the relationship between the \textbf{offender} and the state. 3 This raises a complex question as to whether a criminal offence can properly be described as a dispute.
\item	The circle includes a wide range of participants including not only the \textbf{offender} and the victim but also friends and families, community members, and justice system representatives. The primary distinction between conferencing and circles is that circles do not focus exclusively on the offense and do not limit their solutions to repairing the harm between the victim and the \textbf{offender}.
\item	\textbf{phloem} is a conductive (or vascular) tissue found in plants. \textbf{phloem} carries the products of photosynthesis (sucrose and glucose) from the leaves to other parts of the plant. The corresponding system that circulates water and minerals from the roots is called the \textbf{xylem}.
\item	\textbf{phloem} and \textbf{xylem} are complex tissues that perform transportation of food and water in a plant. They are the vascular tissues of the plant and together form vascular bundles. They work together as a unit to bring about effective transportation of food, nutrients, minerals and water.
\item	\textbf{phloem} and \textbf{xylem} are complex tissues that perform transportation of food and water in a plant. They are the vascular tissues of the plant and together form vascular bundles.
\item	\textbf{phloem} is a conductive (or vascular) tissue found in plants. \textbf{phloem} carries the products of photosynthesis (sucrose and glucose) from the leaves to other parts of the plant.
\item	Unlike \textbf{xylem} (which is composed primarily of dead cells), the \textbf{phloem} is composed of still-living cells that transport sap. The sap is a water-based solution, but rich in sugars made by the photosynthetic areas.
\item	In \textbf{xylem} vessels water travels by bulk flow rather than cell diffusion. In \textbf{phloem}, concentration of organic substance inside a \textbf{phloem} cell (e.g., leaf) creates a diffusion gradient by which water flows into cells and \textbf{phloem} sap moves from source of organic substance to sugar sinks by turgor pressure.
\item	The mechanism by which sugars are transported through the \textbf{phloem}, from sources to sinks, is called pressure flow. At the sources (usually the leaves), sugar molecules are moved into the sieve elements (\textbf{phloem} cells) through active transport.
\item	\textbf{phloem} carries the products of photosynthesis (sucrose and glucose) from the leaves to other parts of the plant. The corresponding system that circulates water and minerals from the roots is called the \textbf{xylem}.
\item	\textbf{xylem} transports water and soluble mineral nutrients from roots to various parts of the plant. It is responsible for replacing water lost through transpiration and photosynthesis. \textbf{phloem} translocates sugars made by photosynthetic areas of plants to storage organs like roots, tubers or bulbs.
\item	At this time the Industrial Workers of the World had a membership of over 100,000 members. In 1913 William Haywood replaced Vincent Saint John as secretary-treasurer of the Industrial Workers of the World. By this time, the IWW had 100,000 members.
\item	This was not true of the Industrial Workers of the World and as a result many of its members were first and second generation immigrants. Several immigrants such as Mary 'Mother' Jones, Hubert Harrison, Carlo Tresca, Arturo Giovannitti and Joe Haaglund Hill became leaders of the organization.
\item	Chinese Immigration and the Chinese Exclusion Acts. In the 1850s, Chinese workers migrated to the United States, first to work in the gold mines, but also to take agricultural jobs, and factory work, especially in the garment industry.
\item	The Rise of Industrial America, 1877-1900. When in 1873 Mark Twain and Charles Dudley Warner entitled their co-authored novel The Gilded Age, they gave the late nineteenth century its popular name. The term reflected the combination of outward wealth and dazzle with inner corruption and poverty.
\item	American objections to Chinese immigration took many forms, and generally stemmed from economic and cultural tensions, as well as ethnic discrimination. Most Chinese laborers who came to the United States did so in order to send money back to China to support their families there.
\item	The rise of industrial America, the dominance of wage labor, and the growth of cities represented perhaps the greatest changes of the period. Few Americans at the end of the Civil War had anticipated the rapid rise of American industry.
\item	The resulting Angell Treaty permitted the United States to restrict, but not completely prohibit, Chinese immigration. In 1882, Congress passed the Chinese Exclusion Act, which, per the terms of the Angell Treaty, suspended the immigration of Chinese laborers (skilled or unskilled) for a period of 10 years.
\item	Industrial Workers of the World. In 1905 representatives of 43 groups who opposed the policies of American Federation of Labour, formed the radical labour organisation, the Industrial Workers of the World (IWW). The IWW's goal was to promote worker solidarity in the revolutionary struggle to overthrow the employing class.
\item	The railroads powered the industrial economy. They consumed the majority of iron and steel produced in the United States before 1890. As late as 1882, steel rails accounted for 90 percent of the steel production in the United States. They were the nations largest consumer of lumber and a major consumer of coal.
\item	This finally resulted in legislation that aimed to limit future immigration of Chinese workers to the United States, and threatened to sour diplomatic relations between the United States and China.
\item	\textbf{costa rica} is known as a prime Eco-tourism destination so visitors are assured of majestic views, amazing destination spots and a temperate climate. These factors assure medical tourists of an excellent vacation experience that is conducive for recovery and relaxation.
\item	Medical Tours \textbf{costa rica}: Medical Tourism Made Easy! No Other Firm Has Helped More Patients. Receive Care Over the Last 15 Years
\item	Medical Tours \textbf{costa rica} difference: At MTCR, our aim is to become your one-stop shop for health care services, so we have put together packages with you, the medical tourist, in mind, offering a wide variety of specialties.
\item	Cost of Medical Treatment in \textbf{costa rica}. The following are cost comparisons between Medical procedures in \textbf{costa rica} and equivalent procedures in the United States: [sources: 1,2]
\item	Common Treatments done by Medical Tourists in \textbf{costa rica}. Known initially for its excellent dental surgery services, medical tourism in \textbf{costa rica} has spread to a variety of other medical procedures, including: General and cosmetic dentistry; Cosmetic surgery; Aesthetic procedures (botox, skin resurfacing etc) Bariatric and Laparoscopic surgery
\item	Medical Tours \textbf{costa rica} office remains within the hospital and the Cook brothers 15 year relationship running the hospitals insurance office and seven years running the international patient department serves you the client very well.
\item	About us. Medical Tours \textbf{costa rica} has helped thousands of patients and are the innovators in medical travel to \textbf{costa rica}. Brad and Bill Cook are visionaries that saw the writing on the wall while running the International insurance office for \textbf{costa rica}s busiest and most respected hospital The Clinica Biblica.
\item	In an era of rising health care costs and decreased medical coverage, the concept of combining surgery with travel has taken off. The last decade has seen a boom in the health tourism sector in \textbf{costa rica}, especially in the area of plastic surgery.
\item	The World Bank ranked \textbf{costa rica} as having the highest life expectancy, at 78.7 years. This figure is the highest amongst all countries in Latin America, and is equivalent to the level in Canada and higher than the United States by a year. Top Hospitals for Medical Tourism in \textbf{costa rica}
\item	Over the last decade, \textbf{costa rica} has evolved from being a mere eco-tourism destination and emerged as a country of choice for foreigners, particularly from United States and Canada. These seek quality healthcare services and surgeries at a much lower price than their home countries.
\item	Color  urine can be a variety of colors, most often shades of yellow, from very pale or colorless to very dark or amber. Unusual or abnormal urine colors can be the result of a disease process, several medications (e.g., multivitamins can turn urine bright yellow), or the result of eating certain foods.

\end{enumerate}

\end{document}